\documentclass[conference]{IEEEtran}
\IEEEoverridecommandlockouts

\usepackage{cite}
\usepackage{amsmath,amssymb,amsfonts}
\usepackage{algorithmic}
\usepackage{graphicx}
\usepackage{textcomp}
\usepackage{lipsum}
\usepackage{float}
\usepackage{xcolor}
\usepackage{parskip}
\usepackage{setspace}
\usepackage{tabularx}
\usepackage[linesnumbered,ruled,vlined]{algorithm2e}
\usepackage{pdflscape}
\usepackage[utf8]{inputenc}
\usepackage{newunicodechar}
\newunicodechar{ℓ}{\mathbb{R}}
\newunicodechar{Ω}{\mathbb{R}}

\usepackage[hyphens]{url} 
\usepackage[colorlinks=true, urlcolor=blue, breaklinks=true]{hyperref} 

\def\BibTeX{{\rm B\kern-.05em{\sc i\kern-.025em b}\kern-.08em
    T\kern-.1667em\lower.7ex\hbox{E}\kern-.125emX}}

\newcommand*{\affaddr}[1]{#1} 
\newcommand*{\affmark}[1][*]{\textsuperscript{#1}}
\newcommand*{\email}[1]{\texttt{#1}}

\newcommand{\linebreakand}{%
  \end{@IEEEauthorhalign}
  \hfill\mbox{}\par
  \mbox{}\hfill\begin{@IEEEauthorhalign}
}

\begin{document}

\title{Generating Alpha: A Hybrid AI-Driven Trading System Integrating Technical Analysis, Machine Learning and Financial Sentiment for Regime-Adaptive Equity Strategies\thanks{This paper presents the full version of a work accepted for publication in the International Conference on Computing Systems and Intelligent Applications (ComSIA 2026), to appear in Springer Lecture Notes in Networks and Systems (LNNS).}}

\author{%
    Varun Narayan Kannan Pillai\affmark[1]
    Akshay Ajith\affmark[1]
    Sumesh K J\affmark[1]
    \\[0.9em]
    
    \affaddr{\affmark[1]Department of Computer Science and Engineering, Amrita School of Computing,\\Amrita Vishwa Vidyapeetham, Amritapuri, India}\\[0.5em]

    \email{%
        varunnarayankannanpillai@gmail.com,
        akshayajith292@gmail.com, 
        sumeshkj@am.amrita.edu
    }\\[-0.8em]
}

\maketitle

\begin{abstract}

The intricate behavior patterns of financial markets are influenced by fundamental, technical, and psychological factors. During times of high volatility and regime shifts causes many traditional strategies like trend-following or mean-reversion to fail. This paper proposes a hybrid AI-based trading strategy that combines (1) trend-following and directional momentum capture via EMA and MACD, (2) detection of price normalization through mean-reversion using RSI and Bollinger Bands, (3) market psychological interpretation through sentiment analysis using FinBERT, (4) signal generation through  machine learning using XGBoost and (5)dynamically adjusting exposure with market regime filtering based on volatility and return environments. The system achieved a final portfolio value of \$235,492.83, yielding a  return of 135.49\% on initial investment over a period of 24 months. The hybrid model outperformed major benchmark indexes like S\&P\,500\textsuperscript{\textregistered} and NASDAQ-100 over the same period showing strong flexibility and lower downside risk with superior profits  validating the use of multi-modal AI in algorithmic trading.

\end{abstract}


\section{Introduction}

AI has come forth to be one of the groundbreaking innovations of the 21st century. This is due to the fact it allows for machines to undertake activities that require human intelligence. AI encompasses a wide array of computational techniques such as machine learning, deep learning, natural language processing (NLP), and reinforcement learning, to mention a few. Because of modern techniques, machines are now able to identify patterns, predict outcomes, and make complex decisions based on vast amounts of data. As the systems of AI progress, they continue to showcase efficacy in areas that require fast paced and high volume learning such as real-time analysis of data, which is an enticing trait to financial markets.\newline
Within the financial services sector, AI has been adopted for other use cases such as risk evaluation, customer service, algorithmic trading, fraud detection, credit scoring, and managing assets. To add, AI is being leveraged more  for tasks such as quantitative modeling and portfolio optimization. AI also gives better forecasting by leveraging high-dimensional historical data, market micro-structure data, and alternative datasets, which include but are not limited to satellite imagery or social sentiment.Now, financial institutions employ different methods of supervised and unsupervised machine learning techniques to identify hidden factors influencing market behavior and use reinforcement learning structures for dynamic portfolio and hedge rebalancing under uncertain conditions.\newline
Artificial intelligence in stock trading has reshaped the industry by offering responsive, sophisticated options that surpass the fixed decision-making frameworks based on rigid criteria. An AI engine may utilize multiple streams of data, both organized, such as prices, volume, volatility, and technical indicators, as well as disorganized like financial news, social media feeds, earnings call transcripts, and social media. Processing both organized and disorganized data enables AI to extract signals and make predictions with greater accuracy. Unlike traditional models, which often use linear assumptions and fail when there is a change in regime, AI models learn and adapt to complex non-linear dependencies across many different market scenarios. More complex models such as XGBoost, LSTMs, and FinBERT, enhance prediction power by detecting temporal patterns. The incorporation of macroeconomic factors, microstructural signals, and sentiment metrics leads to real-time trade signal generation. Position sizing and risk allocation, can be autonomously optimized through precision execution algorithms that sense volatility and sentiment regimes while providing guidance on what to do next. In algorithm-dominated financial markets, decisions made within milliseconds can determine whether or not there is profit to made, which is why this flexibility becomes crucial. \newline
\begin{figure*}[ht]
  \centering
  \includegraphics[width=9cm, height=9cm]{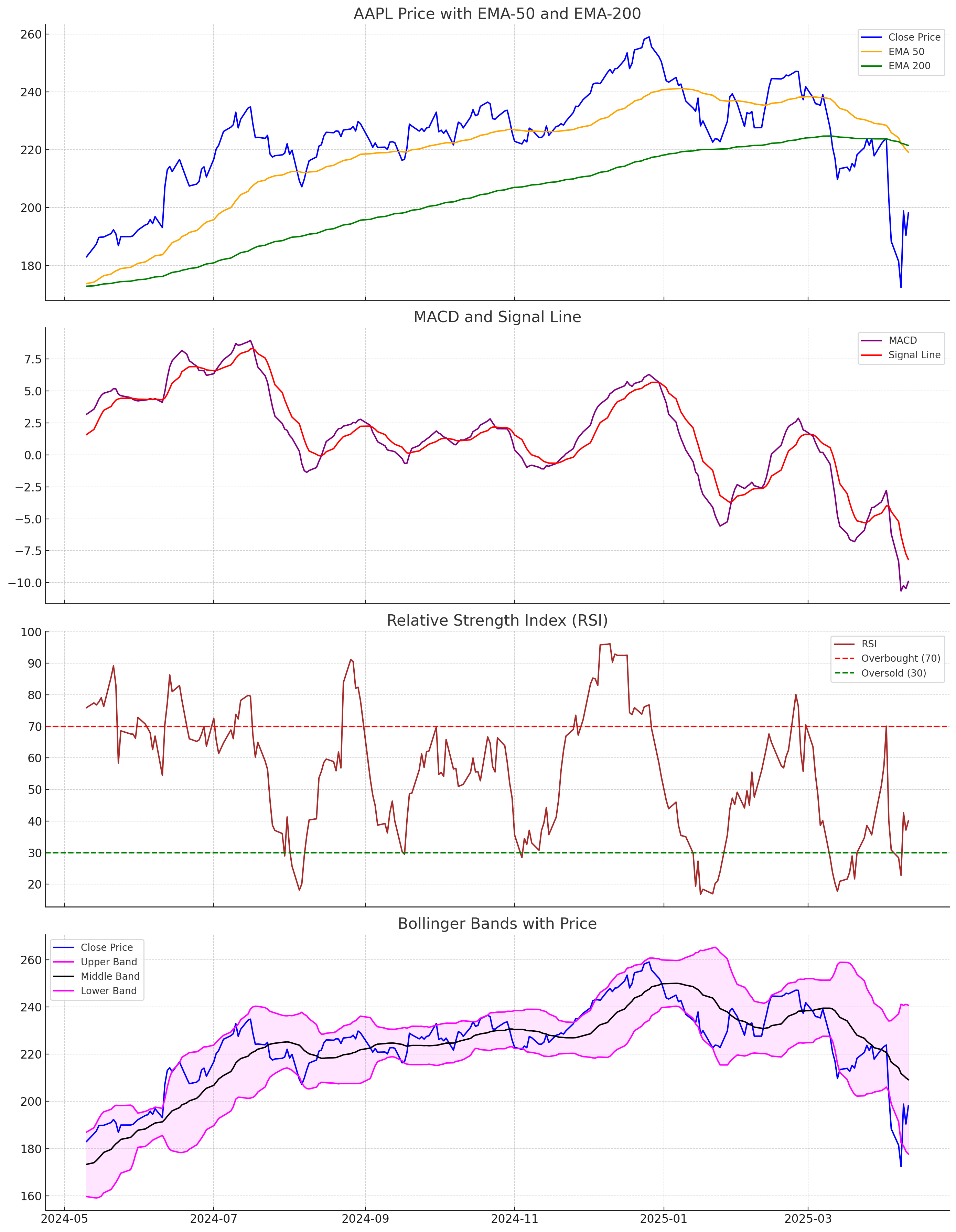} 
  \caption{Apple stock with different technical indicators}
  \label{fig:example1}
\end{figure*}
Financial markets are intricate and ever-changing systems. They are shaped by numerous forces, among which direct and measurable factors like macroeconomic indicators and corporate fundamentals play a significant role. Recently, the financial industry has experienced a shift towards trading systems based on data algorithms powered by AI and machine learning. Even with these advancements, a lot of trading strategies still depend on isolated approaches like using standalone predictive models. These approaches do not succeed in diverse market scenarios.\newline
The objective of this project is to devise a unified risk-aware, robust, and adaptive algorithmic trading system that solves the problem of shifting market conditions through the hybrid approach of combining various alpha-generating signals.\newline
A majority of traditional algorithmic strategies are described as trend-following and mean-reverting. While they can perform well in some scenarios, they often fail to sustain performance when the conditions are changing. For example, trend following strategies depending on EMA or MACD crossovers tend to perform poorly during sideways markets, and mean reversion strategies which depend on RSI or Bollinger Bands tend to provide a lot of false signals in strongly trending markets. The purpose for this research is enhancing the performance of the hybrid AI trading algorithm by integrating machine learning algorithm with  technical indicators and using sentiment analysis control to mitigate losses during negative news cycles.\newline
Let's take an example of a stock like Tesla (TSLA). Cross-over trading strategies may have a macro sentiment bullish switch and tend to strongly buy the stock when the price breaks above the 50-day moving average (50-day EMA) and 200-day moving average (200-day EMA) with MACD indicating bullish momentum as well. However, on the very same day, some breaking news like SEC Investigations or poor quarterly earnings could trigger those drawdown tendencies even with supportive technical reasoning. This model would be blinded to such context if purely technical.\newline
By incorporating FinBERT, a transformer based model trained on financial text, the stock picks can assess sentiment polarity in real-time news headlines and earnings reports. If the average sentiment score is significantly bearish, for example, below -0.70, the system can refrain from placing active trades, thereby acting as a risk shield to core positions on high impact events.\newline
Even with all the advancements made regarding AI in the financial sector, there are still numerous challenges to tackle. Problems arise with sentiment analysis because financial data like news is both unstructured as well as context-dependent, and sentiment analysis requires structure. During market regime shifts, most machine learning models tend to overfit historical data. Most trading strategies are built around the premise of a constant market, rendering them ineffective during transitions, like the switch from bull to bear markets. Risk is usually managed in a blunt way with fixed position limits, neglecting changes in volatility, risk events, or other factors. Furthermore, having multiple indicators generates conflicting signals which adds another layer of complexity to making decisions.\newline
No singular strategy stands out as providing a reliable solution even though a few have attempted. Mean reversion methods that exploit price extremes and utilize RSI or Bollinger bands tend to struggle when breakouts occur. Trend following systems such as EMA and MACD tend to perform poorly during market reversals. Sentiment analysis tools such as FinBERT or VADER are commonly used to label texts, but seldom integrated with trading sensors.\newline
The market regime detection module based on 20-day returns assists in separating bull and bear markets to better tailored strategies. Position sizes are adjusted based on volatility, using the Average True Range (ATR), which provides a more responsive approach to volatility. We utilize Backtrader and simulate trading on 100 diversified S\&P\,500\textsuperscript{\textregistered} stocks for 6 years (4 years of training and 2 years of backtesting) of historical data.\newline
The strategy achieved a better risk-adjusted return, portraying an enhanced Sharpe ratio and lower drawdown reduction. By integrating ML predictions of structured indicators and regime-awareness with sentiment data the encapsulated system showcases clear adaptability and resilience in modern algorithmic trading patterns.
\noindent The main contributions of this work are summarized as follows

\begin{itemize}
\item \textbf{Unified Hybrid AI Driven Trading Architecture}

Our proposed trading trademark includes technical indicators, supervised machine learning, sentiment analysis of financial news and market regime identification in a single decision-making unit. The unified design provides increased robustness to volatility and market regime changes where single strategy designs usually fail.

\item \textbf{Risk Control Mechanism based on Regime and Sentiment Awareness}

The FinBERT used for sentiment analysis of financial news is used as a gate to ensure trades occur only after the sentiment has been analyzed. The use of a low cost regime filter based on returns constrains trades only to favorable market conditions. When working together, both mechanisms provide protection from negative impact of sentiment driven shocks and downside risk associated with changing market regimes.

\item \textbf{Machine Learning using Interpretable Features to Provide Volatility Adaptive Risk Management with Large-Scale Validation}

A directional prediction model using interpretable technical and volatility features and a risk management model utilizing position sizing (using ATR) were developed. The combination of these two models was then empirically validated on a universe of 100 equities of the S\&P\,500\textsuperscript{\textregistered} Index over the time period 2023--2025 and compared to traditional trading strategies and major market indexes.
\end{itemize}

\section{Related Work}

The work by Lo, Mamaysky, and Wang (2000) ``Foundations of Technical Analysis: Computational Algorithms, Statistical Inference, and Empirical Implementation,'' \cite{lo2000foundations} examines technical analysis algorithms through a rigorous statistical lens. The authors use non-parametric kernel regression to evaluate the predictive capability of certain chart patterns like head-and-shoulders and double bottoms. Their results suggest that some patterns do possess considerable forecasting power statistically, especially in more inefficient markets. Although it helped advance the interplay between empirical finance and technical heuristics, their approach is not very flexible in high-frequency or regime-changing contexts. This reason motivates the static learning signal approaches that seek to refine and adapt these technical signals through dynamic learning algorithms.\newline
The 2017 paper titled ``Financial News Predicts Stock Market Volatility Better Than Close Price'' \cite{atkins2018} explained how financial news helps in forecasting future volatility and also discussed its predictive accuracy when compared to close prices. The research shows that models based on text outperform volatility models based on GARCH and historical values. This outperformance is observed when the volatility is realized, implying that traders and other sophisticated market players act on textual information very fast. Hence, they outpace the competition in the market when using intelligent trading strategies. However, the authors do use shallow learning methods and do not employ any form of deep semantics. This gap can be filled with BERT and FinBERT since they use deep learning to form contextual preverbal structures.\newline
The article by Zhengyao Jiang et al. (2017) ``Deep Reinforcement Learning for Portfolio Management'' \cite{jiang2017} suggests a trading agent based on DRL and trained using deep Q-networks (DQN) and policy gradient methods. The agent then learns the optimal allocation weights among a basket of assets in order to maximize cumulative return. They tested their model against traditional strategies and achieved better Sharpe ratios. Though it may be seen as an initial step towards autonomous trading agents, its dependence on synthetic environments and limited incorporation of exogenous variables like macroeconomic indicators, news etc hinders its real-world robustness. Adding other modules such as sentiment filters or macro signals would ideally help the model generalize across market regimes.\newline
In their work, Araci (2019) \cite{araci2019finbert} created a BERT-based transformer model called FinBERT, which is fine-tuned to analyze the sentiment of financial text. FinBERT outperforms generic sentiment models on financial text. FinBERT achieves high accuracy on datasets containing analyst report and financial headlines. Domain-specific corpus finetuning allows the model to grasp subtleties such as negation, hedging, and jargon. Still, with regard to sentiment and multi-ticker contexts, FinBERT can only identify sentiment using named entity recognition or ticker-mapping methods. FinBERT requires significant pre-processing and context handling when dealing with multi-document summarization or real-time news feed processing.\newline
In their paper ``Market Expectations in the Cross-Section of Present Values'' published in the Journal of Finance, Kelly and Pruitt (2013) \cite{kelly2013market} advanced a novel methodology for capturing macroeconomic regime shifts using diffusion index forecasting. The authors perform principal component analysis (PCA) on a rich set of macro indicators to derive latent factors which, through modeling, significantly capture the movement in prices of the assets. These extracted indicators possess predictive ability for excess return forecasts across various asset classes. Although useful, this technique is constrained by the requirement of clean, high-frequency macro data and lacks the granularity in modeling sentiment-driven microstructure effects. Real-time sentiment indicators or regime classifiers could complement some of the existing gaps to enhance strategic allocation models.\newline
Athira et al. \cite{athira2023ml} use a combination of trend analysis and supervised machine learning models to evaluate the behavior of the Indian Stock Market, and demonstrated that while there is value in using classifier models to analyze large amounts of data (in comparison to rule-based models) they will only be successful if there are enough historical feature values to train them upon; and they are based solely on price history. In contrast, Garlapati et al. \cite{garlapati2021prophet} applied two different types of classical time-series forecasting models (ARIMA and Facebook's Prophet model), which provided some reasonable short-term forecasting results; however, neither model is able to account for an abrupt change in regime, nor external shock, and therefore require a more dynamic, and flexible predictive system.\newline
Recently, researchers have begun to develop hybrid and deep-learning based architectures that include both numerical and textual/semantic aspects of the data. Jishag et al. \cite{jishag2020sentiment} analyzed the effects of combining price history with news sentiment from India, and demonstrated empirically that news sentiment enhances predictive accuracy in relation to price-only models. The direction of this research was supported further by Annapoorna et al. \cite{annapoorna2024prophet}, who created an automated workflow for improving the scalability and consistency of the forecasting process utilizing Facebook Prophet pipelines; however, this pipeline does not provide any deeper semantic understanding of the news content. Deep-learning based methods address the lack of semantic understanding of news content more directly. Pradeep et al. \cite{pradeep2024transformer} incorporated BERT embedding into their stock price prediction model, and demonstrated that the use of contextual language representation improved forecasting performance. Previous deep-learning research focused on the National Stock Exchange (NSE) by H. M. et al. \cite{hiransha2018nse} validated the ability of neural network models to predict the market, but showed that the models were prone to overfitting, and did not generalize well between regimes. Together, these works demonstrate the necessity to develop hybrid architectures that unify time-series modeling, deep-semantic embeddings, and regime awareness to provide better, and more adaptable market-prediction frameworks.\newline
In total, these researches are the building blocks for the intelligent trading systems that can be hybrid. Lo et al. (2000) \cite{lo2000foundations} offer a sound theoretical basis for technical analysis in terms of statistics while Atkins et al. (2018) \cite{atkins2018} and Araci (2019) \cite{araci2019finbert} showed how important textual data is in the understanding of market risk and sentiment. Jiang et al. (2017) \cite{jiang2017} demonstrates various applications of reinforcement learning to rebalance portfolios dynamically whereas Kelly \& Pruitt (2013) \cite{kelly2013market} present the use of macroeconomic factors when studying market regimes. It is this nexus between technical analysis, sentiment processing, regime detection, and machine learning that constitutes our approach to designing a hybrid model for trading. This approach aims at integrating price-based signals, news sentiment as well as macro-regime awareness into one adaptive trading engine which has robustness and alpha potential in different market conditions.

\section{Proposed Methodology}

This section outlines the structure, parts, and the automation of our hybrid AI-powered trading strategy which is aimed at generating high-confidence, risk-aware, and regime-sensitive trade signals. This strategy combines multiple domains of knowledge including technical analysis, machine learning (ML), FinBERT sentiment analysis, and filtering the market regime in a holistic approach. It is coded in Python, executed through AWS EC2 for real-time trading with the Alpaca API, and tested with historical data using Backtrader. 
\begin{figure*}[ht]
  \centering
  \includegraphics[width=10.5cm, height=15cm]{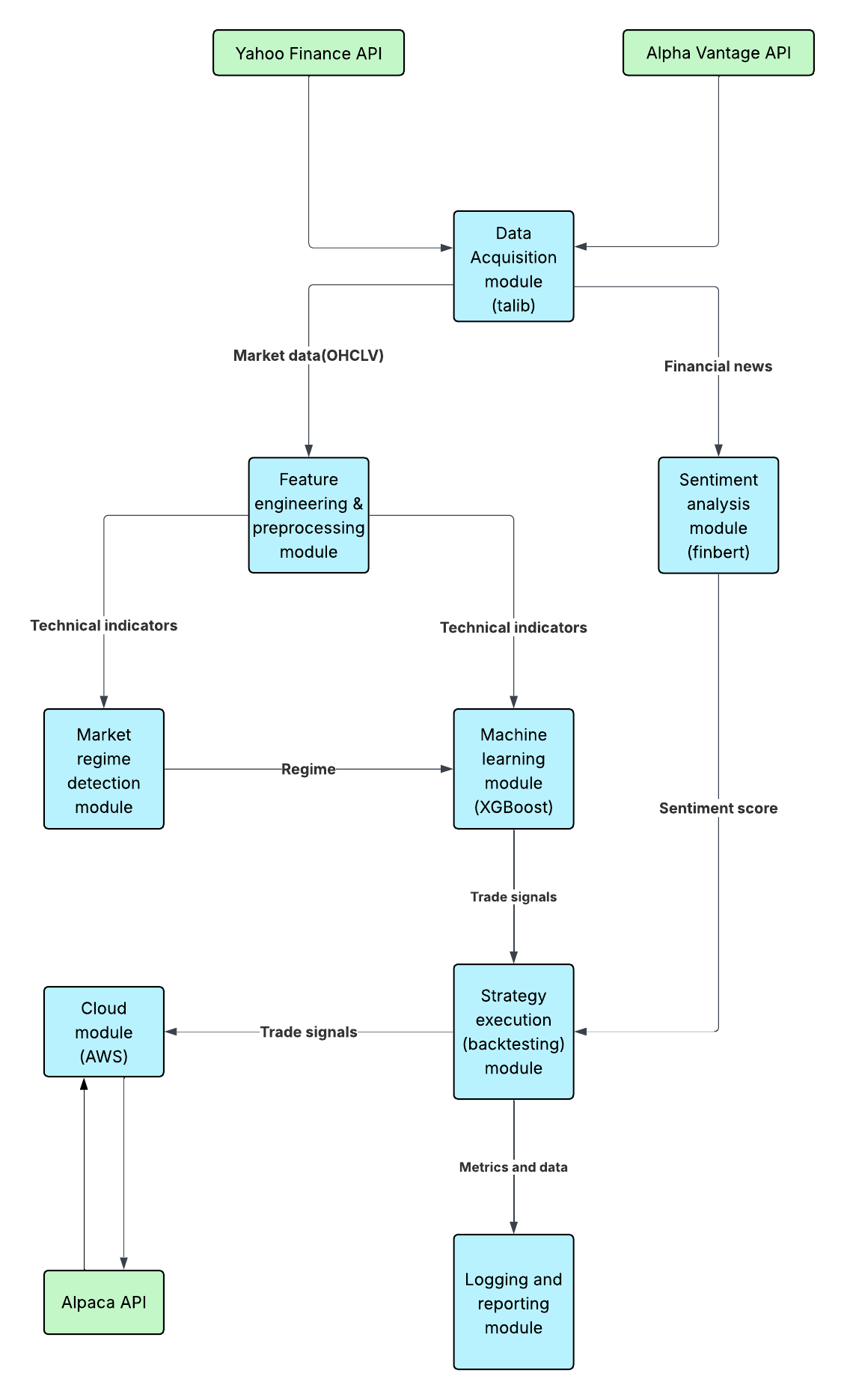} 
  \caption{Proposed Modules}
  \label{fig:example2}
\end{figure*}
\subsection{System Overview}
The system consists of a canonical architecture of eight embedded modules: (1) Data Acquisition, (2) Feature Engineering and Preprocessing, (3) Sentiment Analysis, (4) Machine Learning, (5) Market Regime Detection, (6) Strategy Execution, (7) Logging and Reporting , (8) Cloud Deployment. All modules work together to provide an adaptive, alpha-seeking framework that responds dynamically to market changes and harnesses value from both structured and unstructured data.\newline
Supporting a multi-stock trading universe of over 100 S\&P500\textsuperscript{\textregistered} tickers, the system supports long-only directional trading strategies with daily resolution, albeit restricted by sentiment risk filters and volatility-based position sizing.

\subsection{Data Acquisition}
The system is built on the foundation of data acquisition. We obtain historical OHLCV (Open, High, Low, Close, Volume) data through the yfinance API. Stocks are chosen from a diversified pool which include technology (e.g. AAPL, MSFT, NVDA), financials (e.g. JPM, BAC), and other sectors. The dataset spans from 01/01/2019 to 01/01/2025, amounting to multiple macroeconomic cycles (COVID crash, recovery, inflation shocks) during this period.\newline 
Simultaneously, financial news is retrieved using the Alpha Vantage NEWS SENTIMENT API. For each ticker and each date, headlines along with their corresponding summaries are fetched. To maintain chronological order, only news articles published before 9:30 AM EST on the trading day are considered. All news is structured into CSV files and subsequently stored in a database for advanced processing using the NLP pipeline.

\subsection{Feature Engineering and Preprocessing}

The heart of a predictive model lies in refining and enriching features of untapped financial market data sets. We obtain the following:

\begin{enumerate}

    \item \textbf{EMA 50 AND 200:} Captures short and long-term trends.
    
    \item \textbf{MACD and Signal Line:} Measures momentum via exponential moving average differentials. 

    \item \textbf{RSI (14):} Detects mean-reversion opportunities.

    \item \textbf{Bollinger Bands:} Used for the detection of volatility expansion/contraction. 
    
    \item \textbf{ATR and Volatility (rolling standard deviation):} Used for risk management as well as position sizing.

\end{enumerate}
We also calculate derived features to encode interaction terms depicting non-linear behavior such as the ratio of EMAs (EMA50/EMA200) and the histogram of MACD (MACD – Signal).
Predictive modeling requires accurate and precise features, therefore, all features are standardized and scaled using StandardScaler.
\subsection{Sentiment Analysis}
To bring in behavioral finance, the project implements FinBERT, which is a BERT-based transformer model that was finetuned on financial sentiment analysis. Each news article is run through FinBERT to yield class probabilities for positive, neutral, and negative sentiments.\newline
The per article sentiment scores are aggregated on a daily basis per ticker with a normalized sentiment score: \[
S_t = \frac{1}{N} \sum_{i=1}^{N} \left( P^{(i)}_{\text{positive}} - P^{(i)}_{\text{negative}} \right)
\]
This sentiment score \(St \in [-1,1]\) is used as a filter and not as a signal. Trades are suspended or exited if sentiment breach a threshold of -0.70, which helps defend against downside risk due to adverse news events. FinBERT aids in the identifying of earnings surprises, regulatory captures, guidance cuts, or anything similar even when the prices are lagging.

\begin{figure*}[ht]
  \centering
  \includegraphics[width=10cm, height=3cm]{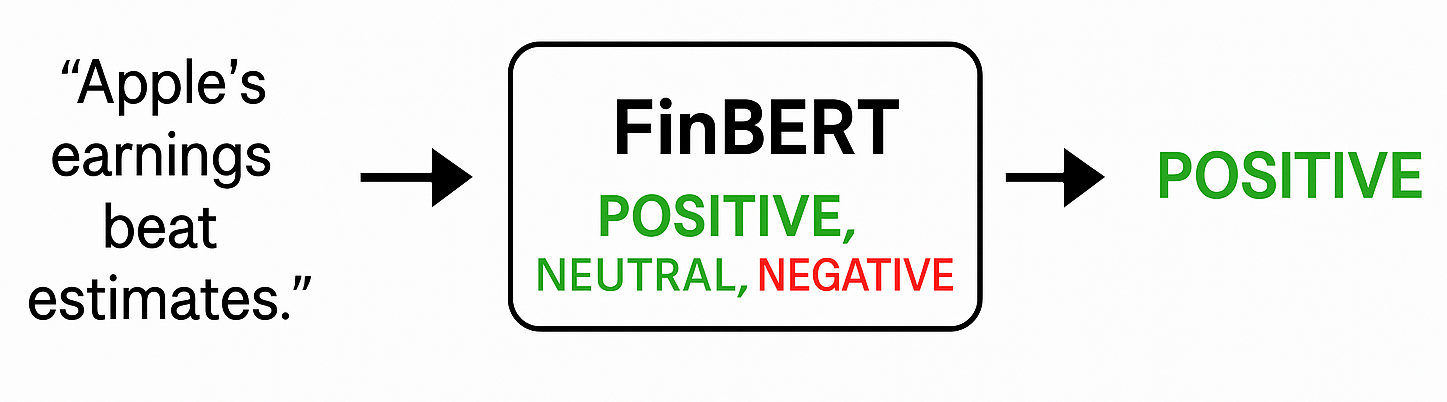} 
  \caption{Analysis of finance news by FinBERT}
  \label{fig:example3}
\end{figure*}

\subsection{Machine Learning}
The primary predictive model is set as an XGBoost classifier that is fitted to predict the next-day return direction. The labels are simply binary, 1 if next-day return $>$ 0, 0 otherwise.\newline
We train the model on concatenated feature vectors of several stocks in order to enhance generalization. The model is provided 10 features as input, where optimization is done using 200 estimators, a depth of 6, and a learning rate of 0.05. This storage contains 70\% of the data for training while 30\% is reserved for testing. The model out-of-sample accuracy is 63\%.\newline
This model is stored and retrieved during live or backtesting sessions. The results are transformed into a probabilistic score which is then altered into binary form. This outcome supplements the trading estimate decision.  
\subsection{Market Regime Detection}
Strategies compliance requires market regime identification. Market regimes are defined as a rolling window average of returns:
\[
R_t = \text{SMA}_{20}(\text{pct\_change}(P_t))
\]
If the output is positive, the regime gets a bullish (1) label, otherwise bearish (-1). This regime identifier works as an on-off switch for enabling or disabling long-sided trades. To address sector specific behaviour, regimes are calculated per stock.
\subsection{Strategy Execution Logic}
The strategy execution engine is implemented using the Backtrader framework. The strategy execution logic is given in Algorithm 1.

A position is opened if no existing position is detected using volatility-based sizing. Volatility-based cash sizing is defined as \[
\text{Shares} = \min\left( \left\lfloor \frac{0.01 \times \text{Cash}}{\text{ATR}} \right\rfloor, \left\lfloor \frac{0.1 \times \text{Cash}}{\text{Price}} \right\rfloor \right)
\]

\begin{algorithm}
\caption{Hybrid Strategy Execution Engine}
\label{alg:hybrid_strategy}
\SetAlgoLined

\KwIn{
Feature vector $X_t$,\\
Technical indicators: EMA$_{50}$, EMA$_{200}$, MACD, Signal, RSI$_{14}$,\\
ATR, Regime $R_t$, Sentiment $S_t$, Price $P_t$, Cash balance,\\
Trained scaler and XGBoost model
}

\KwOut{Trade Action: Buy, Sell, or Hold}

\BlankLine
\textbf{Step 1: Feature Normalization}\\
$X_t^{scaled} \leftarrow \text{scaler.transform}(X_t)$

\BlankLine
\textbf{Step 2: ML Prediction}\\
$\hat{y}_t \leftarrow \text{model.predict}(X_t^{scaled})$

\BlankLine
\textbf{Step 3: Hybrid Score Computation}\\
$\text{score} \leftarrow \hat{y}_t$ \\
\If{$P_t > \text{EMA}_{50}$ \textbf{and} MACD $>$ Signal}{
    $\text{score} \leftarrow \text{score} + 1$
}
\If{RSI$_{14} < 30$}{
    $\text{score} \leftarrow \text{score} + 1$
}

\BlankLine
\textbf{Step 4: Entry Logic}\\
\If{$R_t$ is Bullish \textbf{and} $P_t > \text{EMA}_{200}$ \textbf{and} score $\ge 2$ \textbf{and no open position}}{

    $\text{Risk}_{trade} \leftarrow 0.01 \times \text{Cash}$ \\

    $\text{shares}_{ATR} \leftarrow 
    \left\lfloor \dfrac{\text{Risk}_{trade}}{\text{ATR}} \right\rfloor$ \\

    $\text{shares}_{cap} \leftarrow 
    \left\lfloor \dfrac{0.1 \times \text{Cash}}{P_t} \right\rfloor$ \\

    $\text{shares} \leftarrow \min(\text{shares}_{ATR}, \text{shares}_{cap})$ \\

    \If{$\text{shares} > 0$}{
        \Return \textbf{Buy}$(\text{shares})$
    }
}

\BlankLine
\textbf{Step 5: Exit Logic}\\
\If{$\hat{y}_t = 0$ \textbf{or} {$R_t$ is Bearish} \textbf{or} RSI$_{14} > 70$ \textbf{or}  $S_t < -0.70$ }{
    \Return \textbf{Sell (Full Position)}
}

\BlankLine
\Return \textbf{Hold}

\end{algorithm}
\subsection{Logging And Reporting}
All trades are entered with a timestamp, symbol, action, size, and portfolio value. An ongoing record of average cost and shares per stock is kept. Upon strategy cessation, the following key observations are:
\textit{Final Portfolio Value, Remaining Cash, Total Return (\%), CAGR, Max Drawdown, Sharpe Ratio, Win Ratio, Avg Holding Period.}\newline
These metrics are calculated and logged in portfolio log.txt.
Such metrics are helpful for after the fact evaluation, analysis, and assess against predefined benchmarks.

\subsection{Cloud Deployment with Alpaca on AWS EC2}

For the purposes of live trading, the setup is deployed on an AWS EC2 instance (t2.medium) running Python, Backtrader, and the Alpaca API. A cron job automatically triggers the strategy execution daily prior to the market open. Trades are routed through Alpaca’s paper/live trading interfaces.

Precautions include tokenized encryption, system logs, and trade confirmation emails. News sentiment is obtained through API requests while FinBERT inference is performed using HuggingFace transformers on CPU. This setup provides low latency, high uptime, and reproducible workflows.

\section{Experiments and Results}

\subsection{Experimental Setup}

The development of the hybrid AI-algorithmic trading system was done using Python under object-oriented design principles, where every module is a distinct encapsulated black box. Each component: data capture, feature engineering, predictive modeling, filtering, and trade execution, was validated separately before system integration. This type of modularity allows for an unparalleled level of reusability and decreased maintenance complexity alongside increased system extensibility. System elements such as strategies, data, and even models can be altered freely without issue.

System’s historical simulation backtesting was performed with Backtrader. This powerful event driven Backtest engine simulates market activity in real time ensuring that trading happens only within the bounds of pre-known market information. The model was executed with a starting capital of \$100,000 and set to trade during a two-year backtest period starting January 1, 2023, until January 1, 2025. This period features a rich blend of markets including inflationary tightening cycles with rate increases, earnings volatility as well as post-COVID sectoral rebounds from the COVID-19 pandemic which serve to intensely stress the tested strategy’s resilience.

The yfinance API was used to retrieve historical stock price data, which included OHLCV (Open, High, Low, Close, Volume) values, for 100 large-cap S\&P\,500\textsuperscript{\textregistered}  tickers. The stock pool included companies from different sectors, such as technology, financials, energy, healthcare, and consumer discretionary. A wide array of TA-Lib technical indicators were calculated, including trend-following, mean-reversion, and volatility indicators.  

The Machine Learning module applies XGBoost, an ensemble of gradient boosted decision trees, using a model trained on a multi-ticker dataset created with historical features and forward returns. Custom pipelines were created to retrieve features that included lagged EMA crossovers and histograms of MACD, RSI, Bollinger Band width, volatility bands, and various combinations such as EMA50/EMA200 and MACD-MACD signal which depict the market structure.\newline
To ensure convergence and numerical stability, feature matrices were standardized using Z-score normalization (StandardScaler). Model training and evaluation were performed with an 70-30 time-based split for the training and test datasets to prevent look-ahead bias.\newline
For the purposes of sentiment analysis, it was believed that financial news articles were orderly and could be retrieved via API, such as Alpha Vantage or FinancialModelingPrep. Headlines and summaries were processed using FinBERT, a domain-specific transformer model trained on financial text corpora. Computes of sentiment metrics took place at the ticker-date level and were incorporated into trading logic as a filter preventing entry when sentiment during strongly negative periods worsened (score $<$-0.70).\newline  
A market regime changer delineated bull and bear regimes using a 20-day rolling average return attributed to each trader delineation period as bullish or bearish for trade suppression in macro unfavorable conditions.\newline 
To execute trades, market orders were used not only to simulate realistic fill behavior, which is important for retail brokerage interfaces, for example, Alpaca, but also to execute trades. The amount of capital allocated for each trade was limited to 10 percent. Position sizing adhered to a risk parity approach, where the volatility-based position size determined trade size through the 14-Day Average True Range (ATR).
\begin{itemize}
  \item Total Return
  \item Final Value
  \item Compound Annual Growth Rate (CAGR)
  \item Sharpe Ratio
  \item Maximum Drawdown
  \item Win Ratio
  \item Average Holding Period
\end{itemize}
All primary metrics were stored as structured logs after the simulation, enabling re-calculation and validation of results while maintaining reproducibility standards of empirical science.
\subsection{Experiment 1: Baseline vs. Hybrid Strategy Comparison}

To assess the performance of the hybrid architecture, we evaluated it with a traditional rule-based baseline strategy. The baseline performed trend confirmation using simple moving average (SMA-50/SMA-200) crossovers, and for mean-reversion entries employed RSI (14). There was no adaptive sentiment scaling, adaptive sizing, or any form of sentiment filtering, and the entirety lacked machine learning components.

\begin{table}[ht]
\centering
\scriptsize
\begin{tabular}{|l|r|r|}
\hline
\textbf{Metric} & \textbf{Hybrid Strategy} & \textbf{Baseline Strategy} \\
\hline
Final Portfolio Value & \$235,492.83& \$108,643.27 \\
Market Value of Positions & \$166,042.01 & \$29,184.72 \\
Cash Balance (\$) & \$69,450.82 & \$79,458.55 \\
Total Return (\%) & 135.49\% & 8.64\% \\
CAGR (\%) & 53.46\% & 4.23\% \\
Max Drawdown (\%) & -15.6\% & -19.84\% \\
Sharpe Ratio & 1.68 & 0.48 \\
Win Ratio (\%) & 61.5\% & 53.40\% \\
Avg Holding Period (days) & 5.8 & 11.4 \\
\hline
\end{tabular}
\caption{Comparison of Hybrid and Baseline Trading Strategies}
\label{tab:strategy_comparison}
\end{table}

The hybrid model outperformed the baseline by more than 126\% in total return. The CAGR exceeding 50\% substantiates that the hybrid model not only enhances the magnitude of return but also improves the efficiency of compounding. Sharpe ratio of 1.68 evidences greater performance on risk-adjusted metrics relative to the baseline, exceeding it by 250\%.\newline
 The hybrid model achieved a better win rate suggesting higher precision in entry and exit timing. Sentence filters based on FinBERT prevented entry during news cycles dominated by bearish sentiment, while regime detection limited trades during perpetually bearish phases, which improved risk and profit potential.\newline
The initial model, with all its limitations, had a maximum drawdown of -19.84\% which was almost 28\% more than the hybrid model's drawdown. This was due to overexposure during high volatility or negative sentiment periods.

\subsection{Experiment 2: Benchmark Comparison with Market Indexes}
To benchmark in the hybrid strategy's case, it was measured against US market indexes as of 01/01/2023 to 01/01/2025. The comparison includes:
\begin{itemize}
  \item SPX – Standard S\&P\,500\textsuperscript{\textregistered}, shows market performance as a whole.
  \item NDX - NASDAQ-100, shows growth stocks, especially technology stocks.
  \item DJI – Represents very large capital companies (blue chips).
\end{itemize}

\begin{table}[ht]
\centering
\resizebox{1\linewidth}{!}{%
\begin{tabular}{|l|c|c|c|c|}
\hline
\textbf{Strategy / Index} & \textbf{Final Value (\$)} & \textbf{Return (\%)} & \textbf{CAGR (\%)} \\
\hline
Hybrid Strategy   & \$235,492.83  & 135.49\% & 53.46\%   \\
S\&P 500\textsuperscript{\textregistered} (SPX)    & \$153,187.39  & 53.18\%  & 23.56\%  \\
NASDAQ-100 (NDX)  & \$192,071.58  & 92.07\%  & 36.78\%   \\
Dow Jones (DJI)   & \$128,349.17  & 28.35\%  & 13.27\%  \\
\hline
\end{tabular}
}
\caption{Performance Comparison of Hybrid Strategy vs Major Indices}
\label{tab:strategy_vs_indices}
\end{table}

In order to evaluate the strategy in real-world scenarios, the performance of the hybrid strategy was evaluated against major U.S. market indexes. These were all normalized to a starting capital of \$100,000, which is also the amount used for the comparison checklist.\newline
 The hybrid strategy outperforms all three major indexes by generating significant returns. The NASDAQ-100 trails the hybrid model by over 43\% in total return.\newline
\begin{figure*}[ht]
  \centering
  \includegraphics[height=8cm]{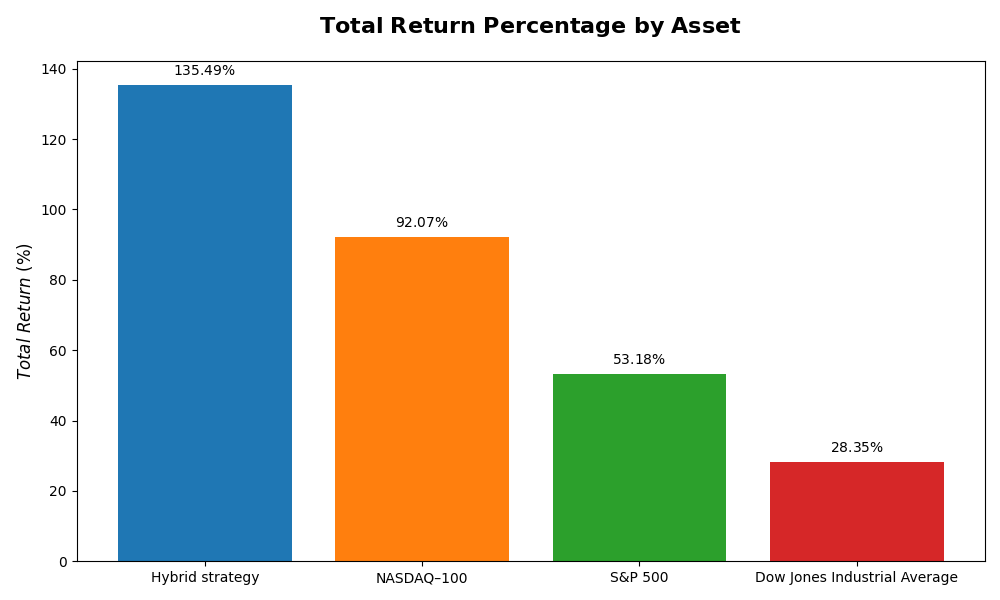} 
  \caption{Comparison with benchmark indexes}
  \label{fig:example4}
\end{figure*}
Compared to index portfolios which are usually static and passive, the main advantage of the hybrid model is its flexibility to respond to macro regime shifts. Almost all passive index portfolios lack these adaptive elements that allow systems to exit early during bearish shocks or avoid exposure altogether when sentiment turns negative.\newline

\section{Conclusion}

This work offers a comprehensive and modular Hybrid AI-Driven Trading Strategy that incorporates classical technical analysis, statistical machine learning, sentiment filtering, and market regime adaptation to manage modern market complexities. It has been designed operating under realistic market conditions and subsequently underwent rigorous testing during a multi-year backtest period from January 2023-January 2025. This period was filled with macroeconomic trends such as bullish runs, interest rate tightening, earnings-driven volatility, and even geopolitical shocks.\newline
As a foundation the strategy utilizes an XGBoost classifier which is fed a myriad of built features including trend indicators like EMA and MACD, mean-reversion signals RSI, along with volatility bands such as Bollinger Bands and ATR. These ticker-centric indicators were computed across 100 diversified S\&P\,500\textsuperscript{\textregistered}  tickers spanning major US sectors and its constituents, allowing the model to learn both cross-sectional and temporal patterns suggestive of short- to medium-term return predictive metrics. To enhance the quantitative core, FinBERT-powered sentiment filtering which transform unstructured financial news into actionable sentiment scores which augment quantitative metrics is implemented. Sentiment scores are employed as gating mechanisms to avoid or exit trades during periods exhibiting heightened negative sentiment thereby adding a behavioral dimension to market signals.\newline
The strategy also includes a separate module that detects the market regime which classifies macro environments as bullish or bearish based off of rolling average returns. Trades are allowed only during bullish regimes which reduces exposure to systemic downturns. Position sizing is scaled with volatility through the use of ATR, while maintaining risk-parity and efficient capital allocation across the portfolio.\newline
Research results show that the provided system outperformed both traditional rule-based systems and passive benchmark indexes. The hybrid strategy’s market value of open positions reached \$166,042.01 and  cash balance of \$69,450.82 having started with \$100,000, resulting in a total return of 135.49\% and CAGR of 53.46\%. These results along with risk-adjusted metrics of Sharpe Ratio 1.68 and Max Drawdown -15.6 indicate the system's effectiveness in sustaining capital during market downswings while maximizing upside during market lulls.
The comparative analysis showed that the hybrid model outperformed S\&P 500\textsuperscript{\textregistered}, NASDAQ-100, and Dow Jones  in returns and capital efficiency. It delivered sector-aligned alpha with the most pronounced effects in technology and consumer discretionary stocks which transcended sectorial borders and demonstrated flexible adaptability in different market regimes—bull, bear, sideways, and high volatility phases.\newline
Noteworthy is the fact that the strategy was implemented on cloud-native architecture using AWS EC2, with integration of Alpaca trading API for live execution, automated email notifications of trades for non-compliance alerting, log alerting auto compliance reporting, and dynamic closed system reporting sentry compliance alerting.\newline
This research reinforces the hypothesis that combining statistical learning, behavioral NLP signals, and awareness of market movements in real-time can improve trading performance significantly, thus validating the thesis. The hybrid model architecture incorporates the strong explanatory power of the classic strategies and models with the enormous predictive capabilities of machine learning, creating an extensible adaptable risk framework for next-gen algorithmic trading.

\section{Future Work}

This system can be further enhanced in multiple ways. To begin with, the scope of sentiment analysis can be augmented to cover real-time unstructured data from social media websites including Twitter, Reddit, StockTwits and YouTube. These platforms target or lead institutional reactions and provide advance alerting mechanisms for crowd behavior or speculative surges. This is further supported by sentiment extraction that can also incorporate transcripts of earnings calls, statements from the CEO and CFO, and documents filed to the SEC such as 10-K and 8-K report. Such documents provide high-signal corporate guidance which are not immediately captured by price movements. Models such as LegalBERT and hybrid transformers can be employed to parse some of these structured disclosures.Additionally, Hidden Markov Models (HMMs) can be introduced for probabilistic regime detection. Exploration of reinforcement learning opens another avenue where agents such as PPO or A2C can dynamically learn trading policies by modifying thresholds and sensitivities relative to market regimes, volatility, and the recorded performance of prior trades. Adopting intraday or minute granularity would also make the strategy suited for high-frequency or swing trading, although this incurs the need for architectural adaptations to enable low-latency inference and execution. Furthermore, enhancing the system to multi-asset domains like ETFs, commodities, and forex will increase diversification of capital and coverage across various markets. This way, transparency can be improved for regulatory frameworks by employing SHAP values or LIME, which would explain transparency enhancement features making institutional reliance possible under regulatory scrutiny. All of these enhancements aim to incorporate intelligence to make the strategy seamlessly agile as an enterprise-level trading algorithm.

\end{document}